\documentclass[11pt]{article}
\usepackage[latin1]{inputenc}
\usepackage[T1]{fontenc}
\usepackage{graphicx}
\usepackage{amsmath}
\usepackage{amssymb}
\usepackage{amsfonts,amsthm}
\usepackage{bbm}
\usepackage{a4}
\usepackage{citesort}
\theoremstyle{plain} 
\newtheorem{Theorem}{Theorem}[section] 
\newtheorem{defi}[Theorem]{Definition}
\newtheorem{Lemma}[Theorem]{Lemma}
\begin{document}
\title{States of Low Energy on Robertson-Walker Spacetimes}
\author{Heiner Olbermann\\
II. Institut für Theoretische Physik\\
Universität Hamburg\\
Luruper Chaussee 149\\
D-22761 Hamburg, Germany\\
E-mail: heiner.olbermann@desy.de}
\maketitle
\begin{abstract}
We construct a new class of physical states of the free Klein-Gordon field in Robertson-Walker spacetimes. This is done by minimizing the expectation value of smeared stress-energy. We get an explicit expression for the state depending on the smearing function. We call it a state of low energy. States of low energy are an improvement of the concept of adiabatic vacua on Robertson-Walker spacetimes. The latter are approximations of the former. It is shown that states of low energy are Hadamard states. 
\end{abstract}

\tableofcontents
\renewcommand{\k}{\ensuremath{\mathbf{k}}}
\newcommand{\Aa}{\ensuremath{\mathcal{A}}}
\newcommand{\Ab}{\ensuremath{\bar{\mathcal{A}}}}
\newcommand{\Ac}{\ensuremath{\hat{\mathcal{A}}}}
\newcommand{\Ad}{\ensuremath{\mathcal{A}^{3}}}
\newcommand{\x}{\ensuremath{\mathbf{x}}}
\newcommand{\y}{\ensuremath{\mathbf{y}}}

\section{Introduction}
In quantum field theory in curved spacetime (QFT in CS), the construction of the algebra of observables is a widely solved problem \cite{Koehler,Hollands,Brunetti}. In principle, the set of physical states is known as well: It is the folium of the Hadamard class. Hadamard states can be defined via the microlocal spectrum condition \cite{Radzikowski}. The set of physical states is then given by their local quasiequivalence class \cite{Verch}.\\ 
The definition of quasifree Hadamard states only involves the singularity structure of the two-point distribution. One does not know how physical properties translate into the smooth part. This is unsatisfactory because there should be Hadamard states of low energy as well as high-temperature Hadamard states and we do not know exactly how to distinguish them. In the present work, such a physical interpretation of an Hadamard state will be given for the first time in Robertson-Walker spacetimes. \\      
The Hadamard concept is rather modern. An older approach to define a class of physical states on Robertson-Walker spacetimes is due to Parker, Lüders and Roberts \cite{Parker,LR}. Parker gave a definition\footnote{The definition given by Parker in \cite{Parker} decribes only approximate states.} of the so-called adabatic vacuum states. This definition was made precise by Lüders and Roberts in \cite{LR}, where they also showed that the adabatic vacuum states (of a certain iteration order) fulfill the principle of local quasiequivalence.
\\
Guided by Parker's ideas and using the techniques of Sobolev wavefront sets, Junker and Schrohe \cite{Junker2} gave a new definition of the class of adiabatic vacua in the language of microlocal analysis and showed that the previous definition was a special case of the new one. They could show that adiabatic states (of a certain order) are locally quasiequivalent to the Hadamard class, thus defining the same set of physical states. Furthermore, they gave a construction procedure for adiabatic vacuum states on globally hyperbolic spacetimes with compact Cauchy surfaces.\\
Nevertheless, the physical interpretation shows that adiabatic vacua should not occur in most physical settings. In an adiabatic vacuum state, the probability of detecting an event of energy $E$ only falls off as a power of $E$ for $E\rightarrow \infty$; in an Hadamard state, the probability decreases faster than any power of $E$, which is the behaviour we know e.g. from all particle or thermal states (see the discussion in \cite{Junker2}).\\
Another drawback of the adiabatic vacua is the unclear physical motivation. Parkers ansatz had been finding states of minimal particle creation. However, in curved spacetimes without asymptotical flatness, a sensible particle interpretation is not available.  
\\
In this work, we present a new approach to define physical states on Robertson-Walker spacetimes and show that these states belong to the Hadamard class. This construction will be physically motivated. The idea is as follows: Based on the results by Lüders and Roberts, we want to find a state with the same symmetry properties as the underlying spacetime that minimizes the energy density measured by an isotropic observer. More precisely, we smear the energy density with a test function supported on the worldline of an isotropic observer and minimize this quantity as a function of the state. This is motivated by a result by Fewster \cite{Fewster}: The renormalized energy density, smeared along a timelike curve, is bounded from below as a function of the state (where only Hadamard states are considered).
\\
The present work is organised as follows: In section \ref{LRrev}, we review the analysis of homogeneous isotropic quasifree pure states on Robertson-Walker spacetimes carried out in \cite{LR}.
In section \ref{slodef}, we show that the above procedure gives indeed a well defined state. We call the resulting state a state of low energy (associated to the smearing function $f$). Applying this result, we show that adiabatic vacuum states are states of low energy in the adiabatic limit. 
The proof that states of low energy satisfy the Hadamard property will be given in section \ref{hadproof}.

\section{Homogeneous Isotropic Quasifree Pure States on Robertson-Walker Spacetimes}
\label{LRrev}
\subsection{Quantum Field Theory in Curved Spacetime}
Robertson-Walker spacetimes have topological structure $M=\mathbb{R}\times \Sigma$ where $\Sigma$ is a Cauchy surface of $M$. For the metric $g_{ab}$ on $M$, we have $g_{ab}=dt²-a(t)²h_{ab}$, where $h_{ab}$ is the pullback of $g_{ab}$ under the embedding map $\iota:\Sigma\rightarrow M$.\\
In this work, we examine the free scalar field in Robertson-Walker spacetimes. We adopt the viewpoint of algebraic quantum field theory (see \cite{Dimock,KayWald}).\\
We start with the set of real-valued test functions on $M$, $\mathcal{D}_\mathbbm{R}(M)=:\Gamma$. There are unique operators $E^\pm:\mathcal{D}(M)\rightarrow\mathcal{E}(M)$ such that (see \cite{Baer})
\begin{eqnarray}
(\square_g+m^2)E^{\pm}=E^{\pm}( \square_g+m^2)=\mathbbm{1}\\
\text{supp } E^{\pm}f\subset J^{\pm}(\text{supp }f).
\end{eqnarray}
$E^\pm$ are the retarded and advanced fundamental solution to the Klein-Gordon operator on $M$. We set $E:=E^+-E^-$ and define an Hermitian form $\gamma$ by
\begin{equation}
\gamma(f,f')=-i\int_M d^4x\sqrt{-g} \left(fE\,f'\right) ,
\end{equation}
where $g$ is the determinant of the metric on $M$. $(\Gamma,\gamma)$ is a symplectic space. The Weyl algebra over $(\Gamma,\gamma)$, $\mathcal{A}$, is generated\footnote{For a more precise statement, see \cite{KayWald}.} by the symbols 
$W(f),f\in\Gamma$ with multiplication law
\begin{equation}
W(f)W(f')=\exp\left(-\frac{1}{2}\gamma(f,f')\right)W(f+f'),
\end{equation}
and star operation
\begin{equation}
W(f)^*=W(-f).
\end{equation}
The symbols $W(f)$ can be understood as exponentials of smeared field operators $\phi(f)$ in a CCR-Algebra, 
\begin{equation}
\label{equivCCR}
W(f)=\exp\left(-i\phi(f)\right).
\end{equation}
To be more precise, a representation of the CCR-Algebra in a normed space induces a representation of $\mathcal{A}$ via equation \eqref{equivCCR} and vice versa. We take the Weyl algebra as our algebra of observables.\\
One can as well start with the symplectic space of initial values to the Cauchy problem, 
\begin{equation}
L=\left\{(f_1,f_2),f_1,f_2\in\mathcal{D}_\mathbbm{R}(\Sigma)\right\}
\end{equation}
with Hermitian form
\begin{equation}
\ell\left((f_1,f_2),(f_1',f_2')\right)=-i\int_\Sigma d^3x\sqrt{h}\left(f_1f_2'-f_1'f_2\right).
\end{equation}
We denote the corresponding Weyl algebra by $\hat{\mathcal{A}}$.\\
${\mathcal{A}}$ and $\hat{\mathcal{A}}$ are isomorphic: Let $\rho_0,\rho_1:C^\infty(M)\rightarrow C^\infty(\Sigma)$ with
\begin{eqnarray}
\rho_0(f)=f|_\Sigma\notag\\
\rho_1(f)=\partial_\Sigma f,
\end{eqnarray}
then an isomorphism $\varsigma:\mathcal{A}\rightarrow\hat{\mathcal{A}}$ is given by
\begin{equation}
\label{eqn:AAhut}
\varsigma({W}(f)) =\hat{W}\left((\rho_0Ef,\rho_1Ef)\right).
\end{equation}
See \cite{KayWald} for a proof.
\subsection{Quantum Field Theory in Robertson-Walker Spacetimes}
Robertson-Walker spacetimes are spatially homogeneous and isotropic. When searching for states of minimal energy density, it is natural to consider only states that have the same symmetry property, as we would think of states that do not have these symmetry properties as excitations of the homogeneous isotropic states. We only consider quasifree or Gaussian states, as our analysis only involves the two-point distributions of homogeneous isotropic states. Quasifree states are entirely characterized by their two-point distribution. Finally, we only consider pure states. This makes sense because we are looking for a state where the smeared energy density is minimal. Energy density is an additive observable so it will not be minimal in a state that can be written as a linear combination of two other states with positive real coefficients unless the smeared energy density is already minimal in these states.\\
\\
We distinguish three cases, $\Sigma=\Sigma^\epsilon,\epsilon\in\{+1,0,-1\}$, depending on whether $\Sigma$ has positive, vanishing or negative curvature. The index $\epsilon$ will be omitted in statements that are valid for all three cases. $\Sigma^\epsilon$ can be given the structure of a submanifold of $\mathbb{R}^4$:
\begin{eqnarray}
\label{xkoord}
\Sigma^+=\{x\in \mathbb{R}^4: (x^0)²+\sum_{i=1}^3 (x^i)²=1\}\notag\\
\Sigma^0=\{x\in \mathbb{R}^4: x^0=0\}\notag\\
\Sigma^-=\{x\in \mathbb{R}^4: (x^0)²-\sum_{i=1}^3 (x^i)²=1, x^0>0\}
\end{eqnarray}
The isometry group $\mathcal{G}^+$ of $\Sigma^+$ is $SO(4)$. Furthermore we have $\mathcal{G}^0=E(3)$ (the euclidean group) and  $\mathcal{G}^-=SO(1,3)=\mathcal{L}^\uparrow_+(4)$ (the Lorentz group). $\mathcal{G}^\epsilon$ can also be thought of as the isometry group of $M^\epsilon=\mathbb{R}\times \Sigma^\epsilon$, as any isometry $\varrho$ of $\Sigma$ defines an isometry $\bar{\varrho}$ of $M$  via $\bar{\varrho}(t,x)=(t,\varrho (x))$.\\ 

This gives rise to automorphisms of $\mathcal{A}$ resp. $\hat{\mathcal{A}}$:
\begin{equation}
\alpha_{\varrho}(W(f))=W(\varrho f),\alpha_{\varrho}(\phi(f))=\phi(\varrho f),\quad \varrho \in \mathcal{G}, (\varrho f)(x):=f(\varrho^{-1}(x)).
\end{equation}
We consider the set of initial values for the Cauchy problem $L=\{(f,f'),f,f'\in\mathcal{D}(\Sigma)\}$. We want to determine the form of two-point distributions on $\hat{\mathcal{A}}$, the Weyl algebra over $L$.\\
A quasifree state $\omega$ with two-point distribution $\omega_2$ is homogeneous and isotropic if and only if
\begin{equation}
\omega_2(f,f')=\omega_2(\varrho f,\varrho f'),\varrho\in \mathcal{G};f,f'\in\mathcal{D}(M).
\end{equation}
 
$\mathcal{G}$ has a representation $U\oplus U$ in $L$ with
\begin{equation}
\label{eqn:dst1}
U(\varrho)f=f\circ \varrho^{-1}.
\end{equation}
The two-point distribution is a bilinear form on $L$.
Now, we would like to think of the two-point distribution as a bounded operator in some Hilbert space. For this reason, we equip $L$ with a Hilbert space structure. A simple choice is $L^2(\Sigma)\oplus L^2(\Sigma)$ where the measure on $\Sigma$ is induced by the metric $h_{ab}$. $L$ is dense in $L^2(\Sigma)\oplus L^2(\Sigma)$. The scalar product is denoted by $\langle \cdot,\cdot\rangle$.\\ 
In order to represent a larger class of two-point distributions as bounded operators in a Hilbert space, we allow a more general structure $H_\nu$ with scalar product
\begin{equation}
\langle F,F'\rangle_\nu=\langle F,(-\Delta+m^2)^{2\nu}F'\rangle,
\end{equation}
where $F=(f_1,f_2),F'=(f_1',f_2')$ and the action of the Laplacian is defined componentwise. Now, we are ready to impose our continuity condition on the two-point distribution: There has to exist a $\nu\in \mathbbm{N}$ such that $\omega_2$ is a continious bilinear form in $H_\nu$. By the Riesz representation theorem, this gives us a bounded linear operator $\tilde\omega:H_\nu\rightarrow H_\nu$ with  
\begin{equation}
\langle F,\tilde{\omega}F'\rangle_\nu=\omega_2(F,F').
\end{equation}
$\mathcal{G}$ has a unitary representation in $H_\nu$ via \eqref{eqn:dst1}. $\omega_2$ will be the two-point distribution of a homogeneous and isotropic state if and only if $\tilde{\omega}$ is in the commutant of this representation. For the rest of the discussion, we return to $L^2(\Sigma)\oplus L^2(\Sigma)$ using a unitary transformation\footnote{The only reason for the definition of $H_\nu$ was giving a better suited continuity condition on $\omega_2$. }
\begin{eqnarray}
V: & H_\nu\rightarrow L^2(\Sigma)\oplus L^2(\Sigma)\notag\\
   & (f,g)\mapsto (-\Delta+m^2)^\nu f\oplus (-\Delta+m^2)^\nu g.
\end{eqnarray}
$V$ intertwines the representation of $\mathcal{G}$ in $H_\nu$ with the representation $U\oplus U$ in $L^2(\Sigma)\oplus L^2(\Sigma)$. It remains the task to compute the commutant of $U$. 
This has been carried out carefully in \cite{LR}, we only present the results. \\
The invariant subspaces of $L²(\Sigma)$ under $U$ are the eigenspaces of the Laplacian on $\Sigma$. There exists an orthonormal basis of eigenfunctions $Y_{\k}$ with eigenvalues $E(k)$, where $k=\sqrt{\k^2}$:  
\begin{eqnarray}
\int d\k \,Y_{\k}(\x)\overline{Y_{\k}(\x')}=\delta(x-x')\\
\int d\x \,Y_{\k}(\x)\overline{Y_{\k'}(\x)}=\delta(k,k')\\
\Delta Y_{\k}(\x)=-E(k)Y_{\k}(\x).
\end{eqnarray}
(In the cases $\epsilon=0,-1$ these are sets of generalised eigenfunctions.) If $\epsilon=0,-1$, the index $\k$ is from $\mathbbm{R}^3$ (equiped with the Lebesgue measure) and if $\epsilon=+1$, $\k$ is from $\mathbbm{N}^3$ (with the counting measure, see \cite{LR}).
A function $f\in L²(\Sigma)$ has a "fourier transform" \footnote{Actually, only if $\epsilon=0$ this is the fourier transform of $f$. In all three cases, there is a fourier inversion formula \cite{LR}.} 
\begin{equation}
\tilde{f}(\k)=\langle Y_{\k},f\rangle_{L²}=\int d\x \,\overline{Y_k(\x)}f(x).
\end{equation}
Its domain is the momentum space associated to $\Sigma^\epsilon$, $\tilde{\Sigma}^\epsilon$.\\ 
A bounded operator  on some $H_\nu,\nu \in\mathbbm{N}$ commuting with $U$  comes down to a multiplication with a polynomially bounded function of $k$ on $\tilde{\Sigma}$.\\
This yields the form of an homogeneous isotropic two-point distribution:
\begin{equation}
\label{eqn:homiso}
\omega_2(F,F')=
\int d\k\sum_{i,j=0}^1\langle\tilde{F}_i(k),\tilde{\omega}_{ij}(k)\tilde{F}_j(k)\rangle_{L²} ,
\end{equation}
where  $\tilde{\omega}_{ij}(k),i,j\in\{0,1\}$  are measurable, polynomially bounded functions on $\tilde{\Sigma}$, and $F=(F_0,F_1),F'=(F_0',F_1')\in L$.\\
\\
Up to now, we have only used the fact that $\omega_2$ is a homogeneous isotropic bidistribution. Furthermore, we want the corresponding functional on $\hat{\mathcal{A}}$ to be state, i.e.\footnote{We define our two-point distribution on $\hat{\mathcal{A}}$ via a relation similar to \eqref{equivCCR} and $\omega_2(F,F')=\omega\left(\phi(F)^*\phi(F')\right)$. For a precise statement, see \cite{LR,KayWald}.}
\begin{eqnarray}
\omega_2(F,F)\geq 0,\,F\in L & \quad  \text{(positivity)}\label{eqn:pos1}\\
\omega_2(F,F')-\omega_2(F',F)=iE_\Sigma(F,F')\notag\\
=i\int_\Sigma d³x\sqrt{h}(F_0F'_1-F_1F'_0),\,F,F'\in L & \quad  \text{(commutation relations)} \label{eqn:kom}
\end{eqnarray}
From equation \eqref{eqn:pos1} we deduce that the matrix $\tilde{\omega}_{ij}(k)$ has to be positive semidefinite for all $k$:
\begin{eqnarray}
\label{eqn:pos2}
\tilde{\omega}_{01}(k)=\overline{\tilde{\omega}_{10}(k)},\quad
\tilde{\omega}_{00}(k)\geq 0,\notag\\
\tilde{\omega}_{00}(k)\tilde{\omega}_{11}(k)-|\tilde{\omega}_{01}(k)|²\geq 0.
\end{eqnarray}
As $\omega$ is pure, the last line of \eqref{eqn:pos2} has to be an equality.\\
In a last step, one shows that from \eqref{eqn:kom} follows
$$\tilde{\omega}_{01}(k)-\tilde{\omega}_{10}(k)=ia³(t_0).$$ 
Combining these statements on the matrix entries $\tilde{\omega}_{ij}(k)$, one has the following
\begin{Theorem}
\label{homisozst}
The quasifree homogeneous isotropic pure states on Robertson-Walker spacetimes are given by \eqref{eqn:homiso} with
\begin{eqnarray}
\label{eqn:pq}
\tilde{\omega}_{11}(k)=a^6(t_0)|q(k)|²,\quad
\tilde{\omega}_{00}(k)=|p(k)|²,\notag\\
\tilde{\omega}_{01}(k)=\overline{\tilde{\omega}_{10}(k)}=-a^3(t_0)q(k)\overline{p(k)},
\end{eqnarray}
where $p,q$ are polynomially bounded functions on $\tilde{\Sigma}$, with 
\begin{equation}
\label{eqn:pq2}
\overline{q(k)}p(k)-q(k)\overline{p(k)}=i.
\end{equation}
\end{Theorem}

Using the isomorphism from equation \eqref{eqn:AAhut}, we now want to determine the two-point distribution for states on $\mathcal{A}$. This will allow us to compute the stress-energy tensor.
With an explicit expression for the fundamental solution of the Klein-Gordon equation $$E:\mathcal{D}(M)\rightarrow\mathcal{E}(M)$$ from \cite{LR}, the derivation is straightforward. We have to make some remarks on the Klein-Gordon operator on $M$ before giving the result.\\
The Klein-Gordon operator on Robertson-Walker spacetimes in coordinates $(t,\vec{x})$ is given by 
\begin{equation}
\square_g+m^2=\frac{\partial^2}{\partial t^2}+3H(t) \frac{\partial}{\partial t}-a^{-2}(t)\Delta_h+m^2,
\end{equation}
where $a(t)$ is the scale parameter, $H(t)=\dot{a}(t)/a(t)$ the Hubble parameter and $\Delta_h$ the Laplacian on the Cauchy surface $\Sigma$.\\
As mentioned in section \ref{LRrev}, there exists an orthonormal system of eigenfunctions $\{Y_{\k}\}$ of the Laplacian,
\begin{equation}
\Delta_h Y_{\k}=-E(k)Y_{\k}.
\end{equation}
The product $T_k(t)Y_{\k}(x)$ is a solution of the Klein-Gordon equation if
\begin{equation}
\label{eqn:T}
\ddot{T}_k+3H\dot T_k+\omega_k^2T_k=0,\quad \omega_k^2=\frac{E(k)}{a^2}+m^2.
\end{equation}

We are now ready to present the form of the two-point distribution on $M$:\\
Let $\hat{\omega}_2$ be the two-point distribution of a state $\hat{\omega}$ on $\hat{\mathcal{A}}$ fulfilling the conditions of theorem \ref{homisozst}. Then the corresponding two-point distribution on $M$ is given by
\begin{equation}
\label{eqn:zweipunkt}
\omega_2(x,x')=\int d\k\, Y_{\k}(\x)\overline{Y_{\k}(\x')}\bar{T}_k(x^{0})T_k(x^{0'}),
\end{equation}
where $T_k$ is a solution of the time part of the Klein-Gordon equation \eqref{eqn:T} with initial conditions \begin{eqnarray}
\label{eqn:incond}
T=T_k(t_0)=q(k)\notag\\
\dot{T}=\dot{T}_k(t_0)=a^{-3}(t_0)p(k),
\end{eqnarray}
where $t=t_0$ defines the Cauchy surface $\Sigma$.
This together with theorem \ref{homisozst} induces a condition on a solution $T_k$ of \eqref{eqn:T}:
\begin{equation}
\label{wronski}
\bar{T_k}\dot{T_k}-\dot{\bar{T_k}}T_k=ia^3.
\end{equation}
This condition is conserved in time. If $T_k$ fulfills equations \eqref{eqn:T} and \eqref{wronski}, it represents a quasifree homogeneous isotropic pure state. This is what will be meant in the following when speaking of "the state $T_k$".
\subsection{Adiabatic Vacuum States}
Parker \cite{Parker} developed explicit expressions for the mode solutions $T_k(t)$. The corresponding quasifree states were named \textit{adiabatic vacua}. However, Parkers mode solutions are only approximate solutions of the differential equation \eqref{eqn:T}, and thus the corresponding states only approximate states. Lüders and Roberts \cite{LR} improved this concept by merely extracting the initial values for the mode solutions from Parkers iteration procedure and subsequently giving the above construction.\\ 
Parkers ansatz is of the WKB-type:
\begin{equation}
\label{eqn:parker}
T_k(x^0)=a(x^0)^{-3/2}\left(2\Omega_k(x^0)\right)^{-1/2}\exp\left(i\int_{t_0}^{x^0}\Omega_k(t')dt'\right).
\end{equation}
Equation \eqref{eqn:parker} automatically satisfies \eqref{wronski}. If expression \eqref{eqn:parker} is a solution of \eqref{eqn:T}, one has
\begin{equation}
\Omega_k²=\omega_k²-\frac{3\dot{a}²}{4a²}-\frac{3\ddot{a}}{2a}+ \frac{3\dot{\Omega}_k²}{4\Omega_k²}- \frac{1\ddot{\Omega}_k}{2\Omega_k}.
\end{equation}
One tries to solve this equation iteratively: 
\begin{eqnarray}
\Omega_k^{(0)}(t)=\omega_k=\sqrt{a^{-2}(t)E(k)+m²}\notag\\
(\Omega_k^{(n+1)})²=\omega_k²-\frac{3\dot{a}²}{4a²}-\frac{3\ddot{a}}{2a}+ \frac{3(\dot{\Omega}_k^{(n)})²}{4(\Omega_k^{(n)})²}- \frac{1\ddot{\Omega}_k^{(n)}}{2\Omega_k^{(n)}}.
\label{Omegait}
\end{eqnarray}
The resulting states with phase functions $\Omega_k^{(n)}(t)$ are called adiabatic vacua of order $n$.\\
Parkers idea was to define modes that  yield the known mode solutions in the case of static spacetimes ($\dot{a}(t)=0$). If one replaces  $a(t)$ by $a(\epsilon t),\epsilon \in\mathbb{R}$ in equation \eqref{eqn:T}, then the difference between \eqref{eqn:parker} and an exact solution should be $O(\epsilon^{2n+1})$ for $\epsilon\rightarrow 0$. Unfortunately, this ansatz is too optimistic as the iteration procedure does not have the necessary convergence properties. Parkers motivation had been finding states of minimal particle creation. We will not pursue this topic any further. For our purposes, the only thing that matters so far is $\Omega_k^{(n)}(t)=\omega_k(t)+O(\epsilon)$.\\

\section{States of Low Energy}
\label{slodef}

Energy density in one spacetime point is not bounded from below as a function of the state \cite{Haag1}. Fewster \cite{Fewster} showed that if energy density is smeared along a timelike curve, this quantity does have a lower bound when only Hadamard states are considered. We take this as a starting point and consider the energy density smeared with a test function with support on the worldline of an isotropic observer in a Robertson-Walker spacetime. In fact, the smearing function has to be the square of a test function; this is a technical detail of Fewster's result. In the class of quasifree homogeneous isotropic pure states, we look for the state with minimal smeared energy density. Obviously, this state will depend on the choice for the smearing test function.\\ 
\\
We give a short sketch of the derivation of our renormalized stress-energy tensor and the main result of \cite{Fewster}.\\
The classical energy density measured by an observer with 4-velocity $u^a$ is 
\begin{equation}
\rho(t)=T_{ab}(\gamma(t))u^a(t)u^b(t),
\end{equation}
where $T_{ab}$ is the classical stress-energy tensor of the Klein-Gordon field,
\begin{equation}
T_{ab}=\nabla_a\phi\nabla_b\phi- \frac{1}{2}g_{ab}\left(\nabla^c\phi\nabla_c\phi+m²\phi²\right).
\end{equation}
Introducing an orthonormal frame $v_\mu^a$ with $v_0^a=u^a$ and performing a point-splitting procedure, we obtain the regularized energy density
\begin{equation}
\rho(x,x')=\frac{1}{2}\left(\sum_{\mu=0}^{3}(v_\mu)^a(x)(v_\mu)^{b'}(x') \nabla_a|_{x}\phi\nabla_{b'}|_{x'}\phi\right)+\frac{1}{2}m\phi(x)\phi(x').
\end{equation}
This yields the regularized (quantum) energy density in a state $\omega$:
\begin{equation}
\label{regenergy}
\langle T^{\text{reg}}\rangle_\omega(x,x')=\frac{1}{2}\sum_{\mu=0}^{3} \left(\left((v_\mu)^a\nabla_a\otimes(v_\mu)^{b'}\nabla_{b'}\right)\omega_2(x,x')\right) +\frac{1}{2}m\omega_2(x,x'),
\end{equation}
This is a bidistribution on $M$. Fewster \cite{Fewster} showed that the pullback of $\langle T^{\text{reg}}\rangle_\omega(x,x')$ to a timelike curve $\gamma$ is a well-defined distribution $\langle T^{\text{reg}}\rangle_\omega(t,t')$ on $\mathbbm{R}^2$. This yields well-defined expressions for differences\footnote{Considering differences, one does not have to care about renormalization issues.} in smeared energy density on $\gamma$:
\begin{eqnarray}
\Delta W & = &  \int dt f(t)^2 \lim_{t'\rightarrow t}\left(\langle T^{\text{reg}}\rangle_\omega(t,t')-\langle T^{\text{reg}}\rangle_{\omega_0}(t,t')\right)\notag\\
\label{sedif}
& = & \int dt f(t)^2 \left(\langle T\rangle_\omega(t)-\langle T\rangle_{\omega_0}(t)\right)
\end{eqnarray}
where $\omega$ and $\omega_0$ are Hadamard states and $f\in\mathcal{D}(\mathbbm{R})$. The main result of \cite{Fewster} is that the above expression is bounded from below as a function of $\omega$ when considering only Hadamard states. The fact that one has to smear with the square of $f$ is a technical detail of Fewster's proof.\\
When looking for states of minimal smeared energy, we would like to have such a lower bound in order to have a well-posed minimization problem. This is our motivation for looking for the state $\omega$ for which \eqref{sedif} is minimal. As we are not only considering Hadamard states but homogeneous states, we cannot be sure the minimization proplem is well-posed. Anyway, our result will justify this ansatz.

\subsection{Minimizing Smeared Energy}
Now one computes expression \eqref{sedif} for an isotropic observer in an homogeneous state. Using equations \eqref{eqn:zweipunkt} and  \eqref{regenergy}, one gets\footnote{The calculation is omitted here for the sake of brevity. For the case $\epsilon=0$, it is trivial. In the case $\epsilon=+1$, one has to perform a partial integration on $\Sigma$. For $\epsilon=-1$, one needs some calculus techniques for the eigenfunctions of the Laplacian $Y_k$ from the appendix of \cite{LR}. The result can already be found in \cite{Parker}.}
\begin{eqnarray}
\label{eqn:TEnergie}
\int dt\, f(t)² \left(\langle{T}\rangle_{\omega}-\langle{T}\rangle_{\omega_0})\right)\notag\\
=\frac{1}{2}\int dt\int d\mathbf{k}\,f(t)² \left( |\dot{T}_k(t)|^2+(a^{-2}(t)E(k)+m^2)|T_k(t)|^2-...\right)\notag\\
=\frac{1}{2}\int dt\int d\mathbf{k}\,f(t)² \left( |\dot{T}_k(t)|^2+\omega_k^2|T_k(t)|^2-...\right)\notag\\
=\int d\k\int dt\,f(t)^2 \left(\rho_k(t)-\rho_{0,k}(t)\right).
\end{eqnarray}
$\omega$ is the quasifree homogeneous isotropic pure state given by $T_k$. The counterterms for $\omega_0$ have been omitted. Formally, we have
$$ \langle{T}\rangle_{\omega}= \int d\mathbf{k}\, \left( |\dot{T}_k(x^0)|^2+\omega_k^2|T_k(x^0)|^2\right).$$
This is independent of the spatial coordinate because of the homogeneity of the considered state. It is not a well-defined expression as it is infinite; however, the integrand is finite. We will minimize it for each mode. Thus, we will not have to care about renormalization issues. Statements like "smeared energy density $\langle{T}\rangle$ is minimal in $\omega$" can always be understood as "smeared renormalized energy density $\langle{T}^{\text{ren}}\rangle$ is minimal in $\omega$".\\ 
If for each $\k$, we find  a $T_k$ such that 
\begin{equation}
\int dt\, f(t)^2 \left( |\dot{T}_k(x^0)|^2+\omega_k^2|T_k(x^0)|^2\right)
\end{equation}
is minimal, we have found the quasifree homogeneous isotropic pure state with minimal smeared energy density. We will call it the state of low energy associated to $f$. \\
So let $\k$ be fixed.\\
The real solutions $T_k(t)$ of equation \eqref{eqn:T} span a two-dimensional real space, complex solutions are therefore from a space of (real) dimension 4. Let $S_k$ be an arbitrary solution of \eqref{eqn:T} satisfiying condition \eqref{wronski}. The generic solution of equation \eqref{eqn:T} can be written as
\begin{equation}
T_k=\lambda S_k+\mu\bar{S}_k,\quad\lambda,\mu\in\mathbb{C}.
\end{equation}
Condition \eqref{wronski} yields
\begin{equation}
\label{eqn:hyperb}
\bar{T}_k\dot{T}_k-\dot{\bar{T}}_kT_k=ia^3\Rightarrow |\lambda|^2-|\mu|^2=1
\end{equation}
and the dimension of the solution space is reduced to 3. Furthermore, if $T_k$ is a solution, then $\exp(i\alpha)T_k,\alpha\in\mathbb{R}$ is as well. Thus, we can choose $\mu$ to be real. This reduces the number of free parameters to 2, the value of $\mu$ and the phase of $\lambda$.\\
We want to minimize
\begin{eqnarray}
W=\int dt\,f(t)^2 \rho_{k,0}(t)\notag\\
=\frac{1}{2}\int dt\,f(t)^2\left(|\lambda \dot{S}_k(t)+\mu \dot{\bar{S}}_k(t)|^2+\omega_k^2|\lambda S_k(t)+\mu \bar{S}_k(t)|^2\right)\notag\\
=\frac{1}{2}\int dt\,f(t)^2 \bigg{(}(|\lambda|^2+|\mu|^2)\left(|\dot{S}_k(t)|^2+\omega_k^2|S_k(t)|^2\right)+\notag\\
+2\Re \left\{\mu\lambda \left(\dot{S}_k(t)^2+\omega_k^2 S_k(t)^2\right)\right\}\bigg{)}\notag\\
\label{eqn:kurzform}
=(2\mu^2+1)c_1+2\mu\Re\left( \lambda c_2\right)
\end{eqnarray}
where
\begin{eqnarray}
c_1=\frac{1}{2}\int dt\,f(t)^2 \left(|\dot{S}_k(t)|^2+\omega_k^2|S_k(t)|^2\right),\label{c1def}\\
c_2=\frac{1}{2}\int dt\,f(t)^2 \left(\dot{S}_k(t)^2+\omega_k^2S_k(t)^2\right)\label{c2def}.
\end{eqnarray}
Given $\mu>0$, $W$ is minimal for $\mbox{Arg }\lambda=-\mbox{Arg }c_2+\pi=\alpha$, such that 
\begin{equation}
\label{eqn:nurmu}
W=(2\mu²+1)c_1-2\mu\sqrt{\mu²+1}|c_2|.
\end{equation}
The minimum can be found by differentiating with respect to $\mu$. This yields an equation of degree 4 for $\mu$. Two of its solutions can be suppressed  because $c_1>|c_2|$; the two remaining solutions differ only by a factor $-1$. The positive solution is given by
\begin{equation}
\label{eqn:mulos}
\mu=\sqrt{\frac{c_1}{2\sqrt{c_1²-|c_2|²}}-\frac{1}{2}}.
\end{equation}
This yields 
\begin{equation}
\label{eqn:lambdalos}
\lambda=\exp\left(i\alpha\right)\sqrt{\frac{c_1}{2\sqrt{c_1²-|c_2|²}}+\frac{1}{2}}.
\end{equation}
We formulate this main result as a theorem:
\begin{Theorem}
\label{znesatz}
Let $\mathcal{A}$ be the Weyl algebra of the free Klein-Gordon field over a Robertson-Walker spacetime $(M,g_{ab})$ and $f(t)\in\mathcal{D}(\mathbbm{R})$. In the set of homogeneous isotropic quasifree states on $\mathcal{A}$, there is a state $\omega_0$ for which the smeared energy density
\begin{equation}
\int dt f(t)² \langle T^{\text{ren}}\rangle_\omega(t)
\end{equation} 
is minimal. This state is given by its two-point distribution \eqref{eqn:zweipunkt} with
\begin{equation}
T_k(t)=\lambda S_k(t)+\mu \bar{S}_k(t),
\end{equation}
where $S_k$ is an arbitrary solution of equation \eqref{eqn:T} fulfilling condition \eqref{wronski}, and $\lambda,\mu$ are given by equations \eqref{eqn:mulos}, \eqref{eqn:lambdalos}, \eqref{c1def} and \eqref{c2def}. We call $\omega_0$ the state of low energy associated to $f(t)$.
\end{Theorem}
Now $T_{k,0}(t)$ is a state of low energy (associated to $f(t)$) if and only if
\begin{equation}
\label{eqn:encond}
\int dt\,f(t)^2 \left(\dot{T}_{k,0}(t)^2+\omega_k^2T_{k,0}(t)^2\right)=0.
\end{equation}
This can be seen by setting $S_k:=T_{k,0}$.
From equations \eqref{eqn:mulos} and \eqref{eqn:lambdalos} we deduce the state of low energy associated to $f(t)$: 
\begin{equation}
\label{eqn:ergebnis}
\tilde{T}_{k,0}= \exp\left(i\alpha\right)\sqrt{\frac{c_1}{2\sqrt{c_1²-|c_2|²}}+\frac{1}{2}}\,T_{k,0} +\sqrt{\frac{c_1}{2\sqrt{c_1²-|c_2|²}}-\frac{1}{2}}\,\bar{T}_{k,0}
\end{equation}
The right-hand side equals $T_{k,0}$ if and only if $c_2=0$ (where we set $\text{Arg }0\equiv 0$).\\
\\
We remark that a state of minimal energy (by which we mean a state that is low energy for any normalized test function $f\in \mathcal{D}(\mathbb{R}),\int fdx=1$) can only exist in static Robertson-Walker spacetimes. As  
\begin{eqnarray}
\frac{d}{dt}\left(\dot{T}_{k,0}(t)^2+\omega_k^2T_{k,0}(t)^2\right)\notag\\
=2\dot{T}_{k,0}(t)\left(\ddot{T}_{k,0}(t)+\omega_k^2T_{k,0}(t)\right)\notag\\
=-6\dot{T}_{k,0}(t)²H(t),
\end{eqnarray}
equation \eqref{eqn:encond} can only be fulfilled for all $f$ if $H=\dot{a}/a=0$, i.e. $a=\text{const.}$ Thus, a state in non-static Robertson-Walker spacetimes cannot be an exact state of minimal energy. If $a=\text{const.}$, the state of minimal energy is well-known: It is the vacuum of the ultrastatic Robertson-Walker spacetime.\\

\subsection{Adiabatic Vacua Are Approximate States of Low Energy}
We apply the concept of states of low energy to the class of adiabatic vacuum states.
We write down Parker's expression for the mode solutions once more: 
\begin{equation}
T_k(t)=\left(2a(t)^3\Omega_k^{(n)}(t)\right)^{-1/2}\exp\left(i\int_{t_0}^{t}\Omega_k^{(n)}(t')dt'\right).
\end{equation}
We have
\begin{equation}
\dot{T}_k=\left(-\frac{3\dot{a}}{2a}-\frac{\dot{\Omega}_k^{(n)}}{2\Omega_k^{(n)}}-i\Omega_k^{(n)}\right) T_k.
\end{equation}
In the adiabatic limit, $a(t)\curvearrowright a(\epsilon t), \epsilon\rightarrow 0$, we have $\dot{a},\dot{\Omega}_k^{(n)}=O(\epsilon)$ and thus 
\begin{equation}
\dot{T}_{k}(t)^2+\omega_k^2T_{k}(t)^2=\left(O(\epsilon)+(-i\Omega_k^{(n)})²+\omega_k^2\right)T_k² =O(\epsilon),
\end{equation}
where we used $\Omega_k^{(n)}=\omega_k+O(\epsilon)$.
Thus, for any test function $f$,
\begin{equation}
\int dt\,f(t)^2 \left(\dot{T}_{k}(t)^2+\omega_k^2T_{k}(t)^2\right)=O(\epsilon).
\end{equation}
We conclude that adiabatic vacuum states are indeed states of minimal energy in the adiabatic limit.

\section{States of Low Energy Are Hadamard States}
\label{hadproof}
In this section, we show that states of low energy possess the Hadamard property. This shows that states of low energy are a set of physically sensible states.\\ 
The idea of the proof is the following: We compare states of low energy and adiabatic vacuum states and show that for large $k$ and large iteration order $n$, the difference between these two converges to zero. This will be used to show that an Hadamard state (which can be understood as an adiabatic vacuum of infinite order) and a state of low energy differ only in the smooth part.

\subsection{Wavefront Sets, Microlocal Spectrum Condition}
The wavefront set of a distribution $u\in\mathcal{D}'(\mathbb{R}^n)$ contains information about the localization and  direction of singularities of $u$ \cite{Hoermander2}. In the context of QFT in CST, this concept was introduced  by Radzikowski \cite{Radzikowski} to characterize Hadamard states. As we do not want to give the original definition of the Hadamard condition \cite{KayWald}, we state Radzikowski's criterion as a definition.
\begin{defi}
Let $M$ be a globally hyperbolic manifold. A quasifree state $\omega$ on the Weyl Algebra $\mathcal{A}$ over $M$ is called an Hadamard state if the wavefront set of its two-point distribution $\omega_2$ satisfies
\begin{eqnarray}
\text{WF}(\omega_2)= & \{(x_1,k_1),(x_2,k_2)\in T^*(M)\backslash \mathbf{0}\times T^*(M)\backslash \mathbf{0}: \notag\\
& (x_1,k_1)\sim(x_2,-k_2),k_1\triangleright 0\}=:C^+.
\end{eqnarray}
\end{defi}
Here, $(x_1,k_1)\sim(x_2,k_2)$ means that there is a null geodesic between $x_1$ and $x_2$ whose derivative in  $x_1$ equals $k_1$ and equals $x_2$ in $k_2$. $k_1\triangleright 0$ means that $k_1$ is in the forward light cone of $T^*_xM$.
\subsection{Proof of the Hadamard Property} 
For our proof, we have to gather some results on adiabatic vacuum states from the work by Lüders und Roberts \cite{LR}.\\
First, we mention a Lemma ensuring the existence of $\Omega^{(n)}(k,x^0):=\Omega^{(n)}_k(x^0)$ from equation \eqref{Omegait} for large $k$:
\begin{Lemma}
\label{Rnlemma} 
Let $I\subset\mathbb{R}$ be a closed interval. For $n\in\mathbb{N}$ there exists a $\chi_n(I)\geq 0$ such that $\Omega^{(n)}$ is positive on
\begin{equation}
\mathcal{R}_n(I):=\{(k,x^0):x^0\in I,k\geq\chi_n(I)\}
\end{equation}
and all its derivatives with respect to $x^0$ are continious on $\mathcal{R}_n(I)$. Furthermore, there are constants $A,B> 0$ such that
\begin{equation}
A(1+k)\leq \Omega^{(n)}_k(x^0)\leq B(1+k),\quad(k,x^0)\in \mathcal{R}_n(I).
\end{equation}
\end{Lemma}
We introduce classes of functions on $\mathcal{R}_n(I)$ with a special asymptotic behaviour for $k\rightarrow\infty$:
\begin{defi}
Let $f(k,t)=f\in C^\infty\left((k_{\text{min}},\infty)\times I\right))$. If for any $m\in\mathbb{N}$, there exists a constant $c_m> 0$ such that 
\begin{equation}
|\partial_t^m f(k,t)|\leq c_mk^n,\quad t\in I,k\geq k_{\text{min}},
\end{equation}
then we write $f\in \mathcal{Q}_n(I)$.
\end{defi}
In particular, if $f\in \mathcal{Q}_n(I)$, we have $f=O(k^n)$.\\
We know from Lemma \ref{Rnlemma} that $\Omega^{(n)}\in\mathcal{Q}_1(I)$. Moreover, one shows that (see \cite{LR})
\begin{equation}
\frac{\Omega^{(n)^2}}{\Omega^{(n-1)^2}}-1=:\epsilon_n\in\mathcal{Q}_{-2n}.
\end{equation}
Next, we need an estimate for the difference between the explicit expression
\begin{equation}
W_k^{(n)}(t):=\left(2a(t)^3\Omega^{(n)}_k(t)\right)^{-1/2}\exp\left(i\int_{t_1}^t dt'\,\Omega^{(n)}_k(t')\right)
\end{equation}
and the exact solution $S_k^{(n)}(t)$ of equation \eqref{eqn:T} with initial conditions
\begin{equation}
S_k^{(n)}(t_0)=W_k^{(n)}(t_0),\quad \dot{S}_k^{(n)}(t_0)=\dot{W}_k^{(n)}(t_0)
\end{equation}
which represents an adiabatic vacuum state. We write
\begin{equation} 
\label{SkWk}
S_k^{(n)}(t)=\alpha^{(n)}_k(t)W_k^{(n)}(t)+\beta^{(n)}_k(t)\bar{W}_k^{(n)}(t).
\end{equation}
One shows that $\alpha$ and $\beta$ satisfy the following integral equation (see \cite{LR} and \cite{Parker}):
\begin{eqnarray}
\alpha(t)=1-i\int_{t_1}^{t} dt'\,R(t')\left[\alpha(t')+\beta(t')\exp\left(-2i\int_{t_1}^{t'}dt''\,\Omega(t'')\right)\right]\notag\\
\beta(t)=i\int_{t_1}^{t} dt'\,R(t')\left[\beta(t')+\alpha(t')\exp\left(2i\int_{t_1}^{t'}dt''\,\Omega(t'')\right)\right],
\label{integralgleichung}
\end{eqnarray}
where we omitted indices (as we will in the following), and
\begin{equation}
\label{Rkernel}
R=\frac{1}{2}\Omega^{-1}\left(\Omega^2-\frac{3\dot{\Omega}^2}{4\Omega^2}+ \frac{\ddot{\Omega}}{2\Omega}-\omega^2+\frac{3\dot{a}^2}{4a^2}+\frac{3\ddot{a}}{2a}\right).
\end{equation}
We have $R^{(n)}=-\frac{1}{2}\Omega^{(n)}\epsilon_{n+1}\in\mathcal{Q}_{-2n-1}(I)$. The asymptotic behaviour of $\alpha$ and $\beta$ is examined with the aid of standard Volterra-Lotka methods (see \cite{LR}, appendix B). We abbreviate equation \eqref{integralgleichung}:
\begin{equation}
\label{vollot}
x=y+Vx,
\end{equation}
with
\begin{equation}
x=\begin{pmatrix}\alpha\\ \beta\end{pmatrix},\quad y=\begin{pmatrix}1\\ 0\end{pmatrix} \text{ und } 
(Vx)(t)=\int_{t_1}^{t}dt'\,V(t')x(t'),
\end{equation}
where $V(t')$ is a $2\times 2$-matrix ist, defined by equations \eqref{integralgleichung} and \eqref{Rkernel}.\\ 
By equations \eqref{integralgleichung} and \eqref{Rkernel}, there is a $C> 0$ such that 
\begin{equation}
|V^{ij}_k(t)|\leq C(1+k)^{-2n-1}.
\end{equation}
One solves equation \eqref{vollot} in a suitable Banach space, where the norm is given by
\begin{equation}
||x||_w:=\max_{i=1,2}\, \sup_{t\in I}\left|\frac{x^i(t)}{w(t)}\right|,
\end{equation}
with
\begin{equation}
w(t):=e^{L|t-t_1|},\quad L\geq 4C.
\end{equation}
We check that $V$ is contracting:
\begin{equation}
||Vx||_w\leq\frac{2C}{L}(1+k)^{-2n-1}||x||_w\leq\frac{1}{2}||x||_w.
\end{equation}
By the Banach fixed point theorem, the unique solution of equation \eqref{vollot} is given by
\begin{equation}
x=\sum_{n=0}^{\infty}V^ny.
\end{equation}
We are interested in the asymptotic behaviour of $\beta$ and $\alpha-1$.
We have
\begin{eqnarray}
\text{sup}_{t\in I}\left|\frac{\beta_k(t)}{w(t)}\right|\leq||x_k-y||_w\leq\sum_{m=0}^{\infty} ||V_k||_w^m||V_ky||_w \notag\\
\leq\frac{4C}{L}(1+k)^{-2n-1}||y||_w\notag\\
\Rightarrow \quad |\beta_k(t)|\leq (1+k)^{-2n-1}||y||_w\sup_{t\in I}w(t)
\end{eqnarray}
The first inequality is valid as well if we replace $\beta_k(t)$ by $\alpha_k(t)-1$. Considering equation \eqref{integralgleichung}, we see that we have similar estimates for the first derivatives of $\alpha,\beta$ with respect to $t$. To conclude, there are constants $C_i>0,i=1,2,3,4$ such that
\begin{eqnarray}
\label{alphabetaestim}
|\alpha^{(n)}_k(t)-1|\leq C_1(1+k)^{-2n-1},\quad |\dot{\alpha}^{(n)}_k(t)|\leq C_2(1+k)^{-2n-1}\notag\\
|\beta^{(n)}_k(t)|\leq C_3(1+k)^{-2n-1},\quad |\dot{\beta}^{(n)}_k(t)|\leq C_4(1+k)^{-2n-1}.
\end{eqnarray}
With equation \eqref{SkWk} we have $S_k^{(n)}(t)=O(k^{-1/2})$ and $\partial_tS_k^{(n)}(t)=O(k^{1/2})$. By iteratively differentiating equation \eqref{eqn:T}, one obtains
\begin{equation}
\label{Sderiv}
\partial_t^jS_k^{(n)}(t)=O(k^{j-1/2}).
\end{equation}
Now let $T_k(t)$ be a state of low energy associated to a test function $f$ whose support is contained in $I$. $T_k(t_0)$ can be obtained from $S^{(n)}_k(t_0)$ by a Bogoliubov transformation
\begin{equation}
\label{bogol1}
T_k(t_0)=\tilde{\alpha}(k)S^{(n)}_k(t_0)+\tilde{\beta}(k)\bar{S}^{(n)}_k(t_0).
\end{equation}
Here, $\tilde{\alpha}(k),\tilde{\beta}(k)$ do not depend on $t$. Indeed, $\mu(k)=\tilde{\beta}(k)$ and
\begin{equation}
\mu(k)=\frac{1}{\sqrt{2}}\left(\frac{1}{\sqrt{1-\frac{|c_2|^2}{c_1^2}}}-1\right)^{1/2}.
\end{equation}
For $|x|<1$, $\frac{1}{\sqrt{1+x}}$  can be developed in a Taylor series,
\begin{equation}
\frac{1}{\sqrt{1-\frac{|c_2|^2}{c_1^2}}}-1= \frac{1}{2}\frac{|c_2|^2}{c_1^2}+\frac{1}{4}\frac{|c_2|^4}{c_1^4}+...
\end{equation}
Thus, we have to examine the asymptotic behaviour of $\frac{|c_2|}{c_1}$.\\
\\
We have
\begin{eqnarray}
\label{grossesumme}
c_1=\int dt\, f(t)^2\left(|\dot{S}(t)|+\omega^2|{S}(t)|\right)\notag\\
=\int dt\, f(t)^2\bigg{(}\left|\frac{d}{dt}\left(\alpha W +\beta\bar{W}\right)\right|^2 +\omega^2\left|\alpha W+\beta\bar{W}\right|\bigg{)}\notag\\
=\int dt\, f(t)^2\bigg{[}\left(|\alpha|^2+|\beta|^2\right) |\dot{W}|^2+\left(|\dot{\alpha}|^2+ |\dot{\beta}|^2+\omega^2\left(|\alpha|^2+|\beta|^2\right)\right)|W|^2\notag\\
+2\Re\bigg{(}\alpha\dot{\bar{\alpha}}\dot{W}\bar{W}+\alpha\bar{\beta}\dot{W}^2+\alpha\dot{\bar{\beta}} \dot{W}W 
\notag\\
+\dot{\alpha}\bar{\beta}W\dot{W}+\dot{\alpha}\dot{\bar{\beta}}W^2+\beta\dot{\bar{\beta}}\dot{\bar{W}}W +\omega^2\alpha\bar{\beta}W^2\bigg{)}\bigg{]}
\end{eqnarray}
Considering these terms one by one, we first remark
\begin{eqnarray}
|W|=(2a^3\Omega)^{-1/2}\in \mathcal{Q}_{-1/2}(I),\notag\\
\exists A_1,B_1>0: A_1(1+k)^{-1/2}\leq |W| \leq B_1(1+k)^{-1/2}\notag\\
|\dot{W}|=\left|3\frac{\dot{a}}{a}+\frac{\dot{\Omega}}{\Omega}+i\Omega\right| (2a^3\Omega)^{-1/2}\in \mathcal{Q}_{1/2}(I),\notag\\
\exists A_2,B_2>0: A_2(1+k)^{1/2}\leq |\dot{W}| \leq B_1(1+k)^{1/2}.
\end{eqnarray}
If, in the above sum, we insert 
\begin{eqnarray}
\alpha^2|W|^2=\left(1+(\alpha+1)(\alpha-1)\right)|W|^2=|W|^2+O(k^{-2n})\notag\\
\alpha^2|\dot{W}|^2=|\dot{W}|^2+O(k^{-2n}),
\end{eqnarray}
the only terms remaining of order larger than $-2n$ in $k$ in \eqref{grossesumme} are $|\dot{W}|^2$ and $\omega^2|W^2|$:
\begin{eqnarray}
c_1=\int dt\, f(t)^2\left(|\dot{W}|^2+\omega^2|W^2|\right)+O(k^{-2n})\notag\\
\Rightarrow \exists \tilde{A},\tilde{B}>0: \tilde{A}(1+k)\leq c_1 \leq\tilde{B}(1+k).
\label{c1analyse}
\end{eqnarray}
The analysis of the asymptotic behaviour of $c_2$ works out similarly:
\begin{eqnarray}
c_2=\int dt\, f(t)^2\left(\dot{W}^2+\omega^2W^2\right)+O(k^{-2n})\notag\\
=\int dt\, f(t)^2\left(|\dot{W}|^2+\omega^2|W|^2\right)\exp\left(2i\int\Omega\right)+O(k^{-2n}).
\label{Vorlemma}
\end{eqnarray}
We need a Lemma to precise our estimate for $c_2$.
\begin{Lemma}
Let $P(k,t)\in\mathcal{Q}_m(I)$. For $M\in\mathbb{N}$, there is a $C_M> 0$ such that 
\begin{equation}
\left|\int dt\, f(t)^2 P(k,t)\exp\left(2i\int_{t_1}^{t}dt'\Omega(t')\right)\right|\leq C_M (1+k)^{-M+m}.
\end{equation}
\end{Lemma}
Proof: Induction by $M$. For $M=0$, we have
\begin{eqnarray}
\left|\int dt\, f(t)^2 P(k,t)\exp\left(2i\int\Omega\right)\right|\notag\\
\leq \int dt\, f(t)^2 |P(k,t)|\leq C_0(1+k)^m.
\end{eqnarray}
Let the hypothesis be valid for $M-1$.\\
For any complex-valued test function $g(t)$, there is a constant $C_g> 0$ such that
\begin{eqnarray}
\left|\int dt\, g(t) P(k,t)\exp\left(2i\int\Omega\right)\right|\notag\\
\leq \int dt\, |g(t)|\, |P(k,t)|\leq C_g(1+k)^m.
\label{hilf7}
\end{eqnarray}
To carry out the induction step, we choose $g(t):=\left(\frac{i}{2}\partial_t\right)^Mf(t)^2$ and replace $P(k,t)$ by $P(k,t)a(t)^M$. We can do this because $P(k,t)a(t)^M\in \mathcal{Q}_{m}(I)$.  We conclude that there is a $C_M>0$ such that
\begin{eqnarray}
C_M(1+k)^m\geq \left|\int dt\, \left(\left(\frac{i}{2}\partial_t\right)^M f(t)^2\right)P(k,t)a(t)^M\exp\left(2i\int\Omega\right)\right|\notag\\
=\left|\int dt\,f(t)^2\left(-\frac{i}{2}\partial_t\right)^M \left(P(k,t)a(t)^M\exp\left(2i\int\Omega\right)\right)\right|\notag\\
=\left|\int dt\,f(t)^2\sum_{j=0}^{M}\binom{M}{j}\left(\left(-\frac{i}{2}\partial_t\right)^j (P(k,t)a(t)^M) \left(-\frac{i}{2}\partial_t\right)^{M-j}\exp\left(2i\int\Omega\right)\right)\right|
\label{hilf8}
\end{eqnarray}
Considering the last line of equation \eqref{hilf8}, we see that for $j\in\mathbb{N}$,
\begin{eqnarray}
\left(-\frac{i}{2}\partial_t\right)^j\exp\left(2i\int\Omega\right)\notag\\
=\left(\Omega^j-\frac{i}{2}\frac{j(j+1)}{2}\dot{\Omega}\Omega^{j-2}+...\right) \exp\left(2i\int\Omega\right)\notag\\
=\left(\Omega^j+O(k^{j-1})\right)\exp\left(2i\int\Omega\right).
\end{eqnarray}
Moreover
\begin{equation}
\left(-\frac{i}{2}\partial_t\right)^j (P(k,t)a(t)^M)\in \mathcal{Q}_{m}(I).
\end{equation}
Now the last line of \eqref{hilf8} reads
\begin{eqnarray}
\label{hilf9}
\left|\int dt\,f(t)^2\left(P(k,t)a(t)^M\Omega^M+O(k^{M-1+m})\right)\exp\left(2i\int\Omega\right)\right|,
\end{eqnarray}
Now we claim  
\begin{equation}
\label{POmega}
P(k,t)a(t)^M\Omega^M= P(k,t)(k^M+O(k^{M-1})). 
\end{equation}
This can be seen with the aid of equation \eqref{Omegait},
\begin{equation}
\left(\Omega^{(n)}_k\right)^2=\omega_k^2+\tilde{P}_0(k,t)=\frac{k^2}{a^2}+P_0(k,t),
\end{equation}
with $P_0,\tilde{P}_0=O(k^0)$.
If we choose $k$ sufficiently large, such that $\frac{k^2}{a^2}>P_0(k,t)$, then $\Omega^{(n)}_k$ can be developed in a Taylor series:
\begin{eqnarray}
\Omega^{(n)}_k=\frac{k}{a}\sqrt{1+\frac{P_0(k,t)a^2}{k^2}}\notag\\ =\frac{k}{a}\left(1+\frac{1}{2}\frac{P_0(k,t)a^2}{k^2}+...\right)=\frac{k}{a}+O(k^{-1}).
\end{eqnarray}
This proofs the claim \eqref{POmega}. If we insert this result into expression \eqref{hilf9}, we obtain
\begin{eqnarray}
k^M\left|\int dt\,f(t)^2 (P(k,t)+k^{-1}\tilde{P}(k,t))\exp\left(2i\int\Omega\right)\right|,
\end{eqnarray}
where $\tilde{P}(k,t)=O(k^m)$. Using the induction hypothesis, we get 
\begin{eqnarray}
C_M(1+k)^m\geq k^M\left|\int dt\,f(t)^2 (P(k,t)+k^{-1}\tilde{P}(k,t))\exp\left(2i\int\Omega\right)\right|\notag\\
\geq k^M\left|\int dt\,f(t)^2 P(k,t)\exp\left(2i\int\Omega\right)\right|\notag\\
-k^M\left|\int dt\,f(t)^2 k^{-1}\tilde{P}(k,t)\exp\left(2i\int\Omega\right)\right|\notag\\
\geq k^M\left|\int dt\,f(t)^2 P(k,t)\exp\left(2i\int\Omega\right)\right|-k^{M-1}C_{M-1}(1+k)^{-M+1+m}\notag\\
\Rightarrow\left|\int dt\,f(t)^2 P(k,t)\exp\left(2i\int\Omega\right)\right|\leq \tilde{C}_M(1+k)^{m-M}.
\end{eqnarray}
\qed
\\
We apply the Lemma to the first term of equation \eqref{Vorlemma}. Here $m=1$. For $M>2n$ the Lemma yields $c_2=O(k^{-2n})$. By equation \eqref{c1analyse}, we obtain the next
\begin{Lemma}
\label{maintheorem}
Let $\omega_n$ be an adiabatic vacuum state of order $n$, represented by a solution $S^{(n)}_k(t)$ of equation \eqref{eqn:T}(see the remark after equation \eqref{wronski}). Let $\omega$ be the state of low energy associated to $f\in\mathcal{D}(\mathbbm{R})$, represented by $T_k(t)$. Then
\begin{equation}
\label{avzne1}
T_k(t)=\tilde{\alpha}(k) S^{(n)}_k(t)+\tilde{\beta}(k)\bar{S}^{(n)}_k(t),
\end{equation}
with 
\begin{equation}
\label{avzne2}
1-\tilde{\alpha}(k),\tilde{\beta}(k)=\mathcal{O}(k^{-2n}).
\end{equation} 
\end{Lemma}

The next step in our proof is combining Lemma \ref{maintheorem} with Lemma 3.3 from \cite{Junker2} (see also Theorem 6.3 from the same paper). 
In order to do so, we need some definitions from the theory of Sobolev wavefront sets (see the appendix of \cite{Junker2} or \cite{Hoermander2}). 
The ($C^\infty$-)wavefront set contains information about the location and direction of a distribution's singularities but not about their degree. This information is contained in the so-called Sobolev wavefront sets.  
\begin{defi}
By $H^s(\mathbb{R}^n),s\in\mathbb{R}$ we denote the Sobolev spaces
\begin{equation}
\left\{u\in\mathcal{S}'(\mathbb{R}^n):||u||_s^2:=\int d^n\xi(1+|\xi|^2)^s|\hat{u}(\xi)|^2<\infty\right\}.
\end{equation}
Equiped with $||\cdot||_s$, $H^s(\mathbb{R}^n)$ is a normed space.
\end{defi}
We proceed by giving the definition of local Sobolev spaces: 
\begin{eqnarray}
H^s_{\text{loc}}(\mathbb{R}^n):=\bigg{\{}u\in\mathcal{D}'(\mathbb{R}^n): \notag\\ 
\int d^n\xi(1+|\xi|^2)^s |\widehat{\varphi u}(\xi)|^2<\infty\quad\forall \varphi\in\mathcal{D}(\mathbb{R}^n)\bigg{\}}.
\end{eqnarray}
This definition can easily be generalised to an arbitratry $C^\infty$-manifold $X$: Let $u\in\mathcal{D}'(X)$.
We say $u\in H^s_{\text{loc}}(\Sigma)$ if for any chart $(U,\kappa_U),U\subset X,\kappa_U\rightarrow\mathbb{R}^n$ of $X$ and for any $\varphi\in\mathcal{D}(U)$, 
\begin{eqnarray}
\int d^n\xi(1+|\xi|^2)^s| \widehat{\kappa_{U*}(\varphi u)}(\xi)|^2<\infty.
\end{eqnarray}
One shows that it is sufficient to verify this criterion for an atlas of $X$ (see \cite{Hoermander1}).\\
We show $C^j(\mathbb{R}^N)\subset H^{s}_{\text{loc}}(\mathbb{R}^N)$ for $j>s+N/2$: 
Let $u\in C^j(\mathbb{R}^N)$, $\varphi\in\mathcal{D}(\mathbb{R}^N),|\alpha|\leq j$. Then
\begin{eqnarray}
\left|\xi^\alpha\widehat{\varphi u}(\xi)\right|=\left|\int d^Nx\, \varphi(x)u(x)(i\partial_x)^{\alpha}e^{-i\xi x}\right|\notag\\
=\left|\int d^Nx\, (i\partial_x)^{\alpha}\left(\varphi(x)u(x)\right)e^{-i\xi x}\right|\leq C_\alpha<\infty\notag\\
\Rightarrow \widehat{\varphi u}=O(\xi^{-j}).
\end{eqnarray}
For $j>s+N/2$ this yields the absolute convergence of the integral $\int d^n\xi|\widehat{\varphi u}(\xi)|^2(1+|\xi|^2)^s$ and therefore $C^j(\mathbb{R}^N)\subset H^{s}_{\text{loc}}(\mathbb{R}^N)$. The same statement follows for a $C^\infty$-manifold $\Sigma$:
\begin{equation}
\label{CHmf}
C^j(\Sigma)\subset H^{s}_{\text{loc}}(\Sigma)\quad\forall s<j-\frac{1}{2}\text{dim }\Sigma.
\end{equation}
We come to the definition of Sobolev wavefront sets. First, let $X\subset\mathbb{R}^n$. A set $\Gamma\subset\mathbb{R}^n\backslash\{0\}$ is called conic if for $t>0$, $(x,\xi)\in\Gamma\Rightarrow (x,t\xi)\in\Gamma$.\\
\begin{defi}
Let $u\in\mathcal{D}'(X)$, $x_0\in X,\xi_0\in\mathbb{R}^n\backslash\{0\},s\in\mathbb{R}$. 
We say $(x_0,\xi_0)$ is not in the $H^s$-wavefront set of $u$ ($(x_0,\xi_0)\not\in\text{WF}^s(u)$) if there exist $\varphi\in\mathcal{D}(X)$ with $\varphi(x_0)\not=0$ and an open conic neighbourhood of $\xi_0$ in $\mathbb{R}^n\backslash\{0\}$, $\Gamma$, such that
\begin{equation}
\int_\Gamma d^n\xi(1+|\xi|^2)^s|\widehat{\varphi u}(\xi)|^2<\infty.
\end{equation}
\end{defi}
The definition of Sobolev wavefront sets on manifolds is carried out in a chart. Let $(\kappa,U)$ be a chart, $\varphi\in\mathcal{D}(U)$ and $\Gamma$ a conic neighbourhood of $d\kappa_{x_0}(\xi_0)$ in $\mathbb{R}^n\backslash\{0\}$. Then $T^*X\ni(x_0,\xi_0)\not\in\text{WF}^s(u)$ if
\begin{equation}
\int_\Gamma d^n\xi(1+|\xi|^2)^s|\widehat{\kappa_*\varphi u}(\xi)|^2<\infty.
\end{equation}
One sees WF$^s(u)=\varnothing\Leftrightarrow u\in H^s_{\text{loc}}(X)$. The $C^\infty$-wavefront set is given by
\begin{equation}
\label{WFabschluss}
\text{WF}(u)=\overline{\cup_{s\in\mathbb{R}}\text{WF}^s(u)}.
\end{equation}\\
For sums of distributions, we have the inclusion
\begin{equation}
\text{WF}^s(u+v)\subseteq \text{WF}^s(u)\cup \text{WF}^s(v).
\end{equation}
\\
We are now ready to cite Lemma 3.3 from \cite{Junker2}.
\begin{Lemma}
\label{L33}
Let $\omega_{H,2}$ and $\omega_{n,2}$ be the two-point distributions of a quasifree Hadamard state and a quasifree adiabatic state of order $n$ on a Robertson-Walker spacetime respectively. Then
\begin{equation}
\text{WF}^s(\omega_{H,2}-\omega_{n,2})=\varnothing\quad \forall s<2n+\frac{3}{2}.
\end{equation}
\end{Lemma}
We will use this Lemma to show our main result:
\begin{Theorem}
\label{znehad}
Let $\omega_2$ be the two-point distribution of a state of low energy $\omega$ (associated to an arbitrary test function $f\in \mathcal{D}(\mathbb{R})$) on the Weyl algebra $\mathcal{A}$ over a Robertson-Walker spacetime $M$. Then
\begin{equation}
\text{WF}(\omega_2)=C^+,
\end{equation}
which means that $\omega$ possesses the Hadamard property.
\end{Theorem}
Proof:\\ 
Let $\omega_{H,2}$ be the two-point distribution of an Hadamard state on $\mathcal{A}$. Then
\begin{eqnarray}
\label{WFvereinigung}
\text{WF}(\omega_2)=\text{WF}(\omega_2-\omega_{H,2}+\omega_{H,2})\notag\\ \subseteq\text{WF}(\omega_2-\omega_{H,2})\cup\text{WF}(\omega_{H,2})= \text{WF}(\omega_2-\omega_{H,2})\cup C^+.
\end{eqnarray}
In the case $\text{WF}(\omega_2-\omega_{H,2})=\varnothing$, the inclusion in \eqref{WFvereinigung} becomes an equality, and $\text{WF}(\omega_2)=C^+$.\\
For the two-point distribution $\omega_{2,n}$ of an adiabatic vacuum of iteration order $n$, we have, using Lemma \ref{L33},
\begin{eqnarray}
\text{WF}^s(\omega_2-\omega_{H,2})\subseteq \text{WF}^s(\omega_2-\omega_{n,2})\cup \text{WF}^s(\omega_{H,2}-\omega_{n,2})\notag\\
= \text{WF}^s(\omega_2-\omega_{n,2})\quad\forall s<2n+\frac{3}{2}.
\end{eqnarray}
For $s<2n-10$, we show $\omega_2-\omega_{2,n}\in H_{\text{loc}}^{s}(M)$, and therefore 
\begin{equation}
\text{WF}^s(\omega_2-\omega_{H,2})=\varnothing\quad \forall s<2n-10.
\end{equation}
(The inequality $s<2n-10$ is not optimal. Anyway, it is sufficient to show that for any $s$, there is a $n$ such that $\omega_2-\omega_{2,n}\in H_{\text{loc}}^{s}(M\times M)$.)\\
Then equalities \eqref{WFabschluss} and \eqref{WFvereinigung} yield the assertion from the theorem.\\
So let $\omega_2,\omega_{n,2}$ be given by solutions $T_k(t),S_k^{(n)}(t)$ of \eqref{eqn:T}. Using \eqref{eqn:zweipunkt}, one has 
\begin{eqnarray}
(\omega_2-\omega_{n,2})(f,f')=\int d^4x\,d^4x'\,d\k \sqrt{g(x)}\sqrt{g(x')}\,\overline{f(x)}f(x')\times\notag\\
\times\left( \bar{T}_k(x^{0})T_k(x^{0'})-\bar{S}^{(n)}_k(x^{0})S^{(n)}_k(x^{0'})\right)Y_{\k}(\x)\overline{Y_{\k}(\x')}.
\end{eqnarray}
We need estimates for the derivatives of $Y_\k(\x)$. For $\epsilon=0$,
\begin{equation}
|\partial_\x^\alpha Y_\k(\x)|=|\partial_\x^\alpha e^{i\k\cdot\x}|\leq (1+k^2)^{|\alpha|/2}.
\end{equation}
For $\epsilon=-1$, we use coordinates
\begin{eqnarray} 
\varphi: & \Sigma^- & \rightarrow  \mathbbm{R}^3\notag\\
& (\sqrt{1+\x^2},\x) & \mapsto \x,\notag 
\end{eqnarray}
and gather from appendix C of \cite{LR} that on a compact set $K\subset\mathbb{R}^3$ 
\begin{equation}
|\partial_\x^\alpha Y_\k(\x)|\leq C_{\alpha,K}(1+k^2)^{|\alpha|/2}.
\end{equation}
As in \eqref{xkoord}, $\Sigma^+$ is considered as the 3-sphere embedded in $\mathbb{R}^4$,  $\Sigma^+=\{x\in\mathbb{R}^4:|x|=1\}$. Again using results from the appendix C of \cite{LR}, we write $Y_\k$ as a function of $x=(x^0,x^1,x^2,x^3)$, whith 
\begin{equation}
\text{sup}_{|x|=1}|\partial_x^\alpha Y_\k(x)|\leq C_\alpha (1+k^2)^{1+|\alpha|/2}.
\end{equation}
In all three cases, we have
\begin{equation}
|\partial_\x^\alpha Y_\k(\x)|\leq C_\alpha (1+k^2)^{1+|\alpha|/2}=O(k^{|\alpha|+2}).
\end{equation}
By equations \eqref{avzne1},\eqref{avzne2} and \eqref{Sderiv}, we have
\begin{equation}
\partial_{(x^0,x^{0'})}^\alpha \left( \bar{T}_k(x^{0})T_k(x^{0'})-\bar{S}^{(n)}_k(x^{0})S^{(n)}_k(x^{0'})\right) =O(k^{|\alpha|-2n-1})
\end{equation}
Therefore, if $4+|\alpha|-2n-1<-3$, the following integral converges absolutely:\footnote{In all of the three cases ($\epsilon=+1,0,-1$), a sufficient criterion for absolute convergence of the integral is an asymptotic behaviour of the integrand as $O(k^{-3-\delta}),\delta>0$ .}
\begin{eqnarray}
\int d^3\k \,\partial_{(x,x')}^\alpha \left[\left( \bar{T}_k(x^{0})T_k(x^{0'})-\bar{S}^{(n)}_k(x^{0})S^{(n)}_k(x^{0'})\right)  Y_{\k}(\x)\overline{Y_{\k}(\x')}\right]\notag\\
=\partial_{(x,x')}^\alpha\left(\omega_2-\omega_{n,2}\right)
\end{eqnarray}
For $j<2n-6$, this implies
\begin{equation}
\omega_2-\omega_{n,2}\in C^j(M\times M).
\end{equation}
Using equation \eqref{CHmf}, we finally obtain  
\begin{equation}
\omega_2-\omega_{n,2}\in  H^s_{\text{loc}}(M\times M)\quad \forall s<2n-10.
\end{equation}
\qed

\section{Summary and Outlook}
\label{summary}

We have given the expression for a quasifree state of the free Klein-Gordon field on Robertson-Walker spacetimes that minimizes the expectation value of the smeared stress-energy tensor, where the smearing was done with the square of an arbitrary test function along the curve of an isotropic observer. This state depends on the smearing function $f$. We called it state of low energy (associated to the smearing function $f$).\\
It was shown that adiabatic vacuum states are states of low energy (for any smearing function) in the adiabatic limit. In static Robertson-Walker spacetimes, the unique state of low energy is the ground state; this means that for any test function the associated state of low energy is the ground state. In non-static Robertson-Walker spacetimes, there cannot exist such a unique state of low energy. \\
We have given the proof that a state of low energy (associated to an arbitrary test function) is an Hadamard state. This has been done by combining results of Lüders and Roberts on the one side and Junker and Schrohe on the other: The higher one chooses the iteration order of an adiabatic state, the closer it comes to being of low energy. This statement expressed in the language of microlocal analysis yields the result.\\
A similar procedure should work out for temperature states: Again, we consider a seperation by modes. Instead of minimizing each mode with respect to smeared energy density, one could try to get a KMS-like behaviour with respect to this operator. One expects to obtain mixed Hadamard states.\\
The concept of states of low energy could also be useful for models of the early universe. Up to now, there has been no precise criterion for choosing the right state of the universe during expansion. States of low energy could be a sensible concept. However, it remains the task to determine the correct smearing function.\\

\centerline{\textbf{Acknowledgements}}
This article is a shortened version of my diploma thesis \cite{DA} which has been supervised by Klaus Fredenhagen. I would like to thank him for important ideas and clarifying discussions. As well, I would like to thank Chris Fewster and Romeo Brunetti for helpful discussions.\\ 
Part of this work has been done at the Erwin Schrödinger International Institute for Mathematical Physics in Vienna.  
\bibliography{slo.bib}

\end{document}